\documentclass[aps,prb,reprint]{revtex4-2}

\usepackage{amsmath,amsfonts,amssymb,graphicx,braket,bbold,mathtools,array}
\usepackage{multirow}
\usepackage{float}
\usepackage[]{units}
\usepackage{hyperref}
\hypersetup{
    colorlinks=true,
    linkcolor=blue,
    filecolor=blue,      
    urlcolor=blue,
    citecolor=blue
}

\usepackage[table]{xcolor}

\usepackage[T1]{fontenc}
\usepackage{tikz}
\usetikzlibrary{calc,patterns,decorations.pathmorphing,decorations.markings, shapes, positioning, shapes.geometric}

\makeatletter
      \def\CT@@do@color{%
      \global\let\CT@do@color\relax
            \@tempdima\wd\z@
            \advance\@tempdima\@tempdimb
            \advance\@tempdima\@tempdimc
    \advance\@tempdimb\tabcolsep
    \advance\@tempdimc\tabcolsep
    \advance\@tempdima2\tabcolsep
            \kern-\@tempdimb
            \leaders\vrule
                    \hskip\@tempdima\@plus  1fill
            \kern-\@tempdimc
            \hskip-\wd\z@ \@plus -1fill }
    \makeatother

\newcommand*{\figref}[2][]{%
  \hyperref[{fig:#2}]{%
   \ref*{fig:#2}%
    \ifx\\#1\\%
    \else #1%
    \fi
  }%
}

\begin{document}
\title{Pauli spin blockade with site-dependent g-tensors and spin-polarized leads}
\author{Philipp M. Mutter}
\email{philipp.mutter@uni-konstanz.de}
\author{Guido Burkard}
\email{guido.burkard@uni-konstanz.de}
\affiliation{Department of Physics, University of Konstanz, D-78457 Konstanz, Germany}

\begin{abstract}
Pauli spin blockade (PSB) in double quantum dots (DQDs) has matured into a prime technique for precise measurements of nanoscale system parameters. In this work we demonstrate that  systems with site-dependent g-tensors and spin-polarized leads allow for a complete characterization of the g-tensors in the dots by magnetotransport experiments alone. Additionally, we show that special polarization configurations can enhance the often elusive magnetotransport signal, rendering the proposed technique robust against noise in the system, and inducing a giant magnetoresistance effect. Finally, we incorporate the effects of the spin-orbit interaction (SOI) and show that in this case the leakage current contains information about the degree of spin polarization in the leads.
\end{abstract}

\maketitle
\section{Introduction}
The search for a scalable quantum computer has seen a steady increase in the complexity of solid-state quantum dot systems and their manipulation \cite{Hanson2007review, Zwanenburg2013, Zhang2019}. As a consequence, it is more important than ever to have as precise a knowledge about system parameters as possible. The Pauli spin blockade (PSB) - or rather the lifting of the blockade - has become an important tool in extracting information from the system under consideration using magnetotransport measurements. At its heart, PSB is the inability of a triplet state formed from one electron in each of the two dots of a DQD to transition to a configuration with both electrons in the right dot. The reason for this blockade is that the only energetically available state in the (0,2) charge configuration is a singlet, and the transition is thus forbidden by spin conservation. The blockade may then be harnessed to read out the spin of the electron or hole via charge sensing \cite{Morello2010, Simmons2011, Buech2013, Watzinger2018, Hendrickx2020, Hendrickx2020b}. Lifting of the blockade may occur by various mechanisms that influence the spin of the particle such as interaction with the nuclear spin bath \cite{Jouravlev2006}, the spin-orbit interaction \cite{Danon2009, Nadj-Perge2010a,Froning2021} and combinations of the above  in systems with a valley degree of freedom and disorder \cite{Palyi2009, Palyi2010}.  Site-dependent g-tensors, which occur due to unavoidable imperfections in quantum dot growth or engineering \cite{Kiselev1998, Nenashev2003, Nakaoka2004, Nakaoka2005, Oliveira2008, Aleshkin2008}, may also lift the PSB, and the resulting leakage current carries information about the spin-orbit vector of the system and the g-tensor components in the dots \cite{Mutter2020}. Conversely, knowledge of the g-tensors can be used to precisely measure magnetic fields in the context of magnetometry \cite{Szechenyi2015, Szechenyi2017}.

There has been broad interest in spin-polarized leads coupled to QDs in recent years, e.g.,  for the purpose of qubit initialization and read-out \cite{Sachrajda2001, Bird2003, Leggett2004,Weymann2019}. Still, the vast majority of investigations on spin blockade lifting in DQDs assumes that the leads are unpolarized in the spin of the particles. In this paper we drop this assumption and consider the case of arbitrarily spin-polarized leads. Previous studies containing DQD systems coupled to spin-polarized leads looked at fixed polarizations \cite{Tanaka2004, Fransson2006, Weymann2007} or characterized the transport by exploring the electric conductance and transmission probabilities \cite{Martinek2003, Yuan2007, Tao2008, Hornberger2008, Dias2013}. In contrast, we focus on the leakage current and by deriving effective lead-dot tunneling rates show that the form of the current is sensitive to the degree of spin polarization in the leads.

We start by introducing the model and formalism of spin-polarized leads in Sec.~\ref{sec:spin-polarized_leads}. In Sec.~\ref{sec:leakage_current} we proceed by investigating different polarization configurations and demonstrate that there exists a configuration where the maximum of the leakage current contains information about the g-tensors in the dots and another at which the g-tensor resonance found in Ref.~\cite{Mutter2020} is amplified. These results allow for a full determination of the g-tensor components in the two dots from magnetotransport measurements alone. Moving towards a more faithful description of many condensed matter systems, we take into account the SOI in Sec.~\ref{sec:SOI} and show that information about the spin polarizations in the leads may be extracted from  the leakage current. Finally, Sec.~\ref{sec:conclusion} provides a conclusion.
\section{Spin-polarized leads}
\label{sec:spin-polarized_leads}
\begin{figure}
		\includegraphics[width=0.85\linewidth]{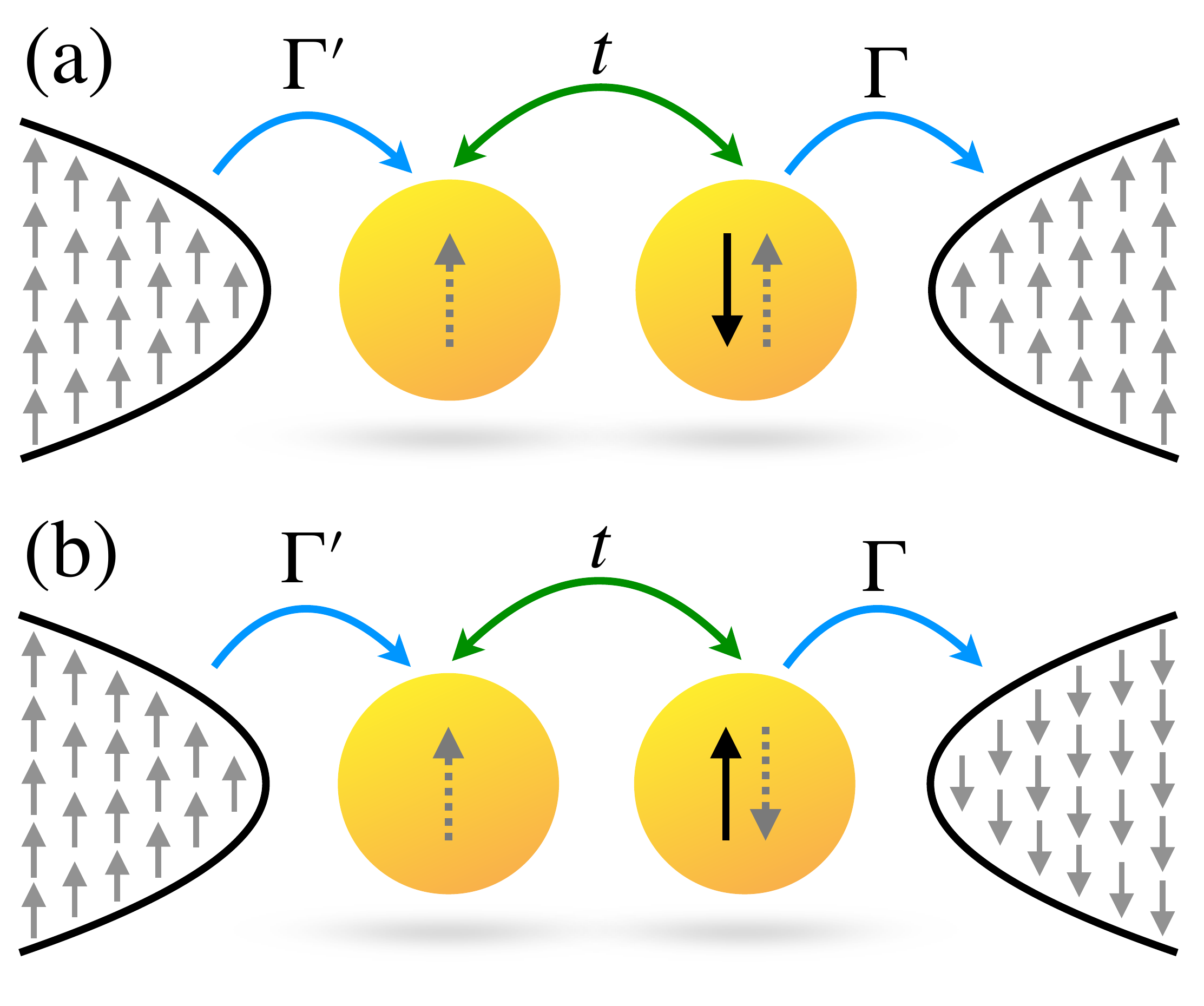}
	\caption{Basic setup for PSB with completely polarized leads. When a bias voltage is applied, electrons may tunnel from the left lead to the left dot with rate $\Gamma'$ and from the right dot to the right lead with rate $\Gamma$ (blue arrows), thereby creating a charge cycle. Once in the DQD, the states in the (1,1) charge configuration must transition to the singlet in the (0,2) charge configuration aided by the inter-dot tunnel coupling $t$ (green arrow). A characteristic feature of completely polarized leads is that one spin making up this singlet is unable to leave the system (solid black arrows). (a) shows one case of complete and parallel polarizations in the leads, $p_L p_R =1$, while (b) shows one case of complete and anti-parallel polarizations, $p_L p_R = -1$.}
	\label{fig:polarized_leads_sketch}
	\end{figure}

The common set-up for PSB is a DQD connected to leads which is tuned to the (0,1)-(1,1)-(0,2) triple point in the charge stability diagram (Fig.~\ref{fig:polarized_leads_sketch}).
Consequently, the relevant two-particle states are
	\begin{align}
	\label{eq:11states}
	\begin{split}
		 &\vert S_{02} \rangle = \frac{1}{\sqrt{2}}\left( \vert \uparrow_R   \downarrow_R \rangle - \vert \downarrow_R  \uparrow_R \rangle \right), \\
		&\vert S \rangle = \frac{1}{2}\left( \vert \uparrow_L  \downarrow_R \rangle - \vert \downarrow_L   \uparrow_R \rangle + \vert \uparrow_R   \downarrow_L \rangle - \vert \downarrow_R   \uparrow_L \rangle \right),  \\
		&\vert T_0 \rangle =  \frac{1}{2}\left( \vert \uparrow_L  \downarrow_R \rangle + \vert \downarrow_L   \uparrow_R \rangle - \vert \uparrow_R   \downarrow_L \rangle - \vert \downarrow_R   \uparrow_L \rangle \right),  \\
		&\vert T_+ \rangle = \frac{1}{\sqrt{2}}\left( \vert \uparrow_L   \uparrow_R \rangle - \vert \uparrow_R   \uparrow_L \rangle \right), \\
		&\vert T_- \rangle = \frac{1}{\sqrt{2}}\left( \vert \downarrow_L   \downarrow_R \rangle - \vert \downarrow_R   \downarrow_L \rangle \right),
		\end{split}
	\end{align}
where $\vert s_d \rangle$ labels a single-particle spin state $s \in \lbrace \uparrow, \downarrow \rbrace$ in dot $d \in \lbrace L,R \rbrace $. The state $\vert S_{02} \rangle$ is the singlet in the (0,2) charge configuration, while $\vert S \rangle$ is the singlet and $\vert T_{0,\pm} \rangle$ are the triplets in the (1,1) configuration. We assume the inter-dot detuning $\epsilon$ to be of the order of the charging energy $U$ and much larger than the tunnel coupling $t$, $t \ll U \sim \epsilon$, such that the (2,0) singlet is energetically well separated and may be neglected. In this regime the basic Hamiltonian describing a tunnel-coupled DQD with site dependent g-tensors in an external magnetic field $\mathbf{B}$ is given by $H = H_0 + H_Z$ with
	\begin{align} \label{eq:H_DQD}
		&H_0 = \Delta \vert S_{02} \rangle \langle S_{02} \vert + t  \left( \vert S \rangle \langle S_{02} \vert + \vert S_{02} \rangle \langle S \vert \right) \\ 
		&H_Z = \sum_d \boldsymbol{\mathcal{B}}^d \cdot \boldsymbol{S}^d,
	\end{align}
where $\Delta$ is the detuning between the (1,1) and (0,2) charge configurations, $t$ is the tunnel matrix element and $\mathcal{B}^d_i = \sum_i g_{ij}^d B_j$ is the effective magnetic field in dot $d \in \lbrace L,R \rbrace$ caused by the anisotropy of the g-tensors. Such anisotropy is observed, e.g., in hole systems in Germanium \cite{Hofmann2019assessing}. Throughout this paper we work in units where $\hbar = \mu_B =1$ such that rates and magnetic fields are measured in units of energy.

When a bias voltage is applied electrons may tunnel to and from the leads, thereby creating a charge cycle. The rate of electrons entering the system by tunneling from the left lead to the left dot is given by $\Gamma'$, while the rate of electrons leaving the DQD system by tunneling from the right dot to the right lead is $\Gamma$. Both these processes are assumed to be spin-conserving, i.e., there is no spin-flip tunneling to and from the leads. In Appendix~\ref{appx:dot-lead_spin-flip_tunneling} we show that spin-flip processes can be incorporated naturally into our model if one wishes to lift this assumption. We work in the common PSB bottleneck setup, $\Gamma' \gg \Gamma$, such that the left dot is effectively refilled with the rate $\Gamma$, i.e., as soon as the second electron leaves the right dot, and for the case of unpolarized leads each of the four states in the (1,1) configuration is refilled with a rate $\Gamma/4$.

To describe the case of non-trivial spin-polarizations, we define the degree of spin polarization (DSP) in a given lead containing spin-$1/2$ particles to be 
	\begin{align}
	\label{eq:spin-polarization}
		p = \frac{2}{N} \sum_{i=1}^N \langle S_z^{i} \rangle \in [ -1,1],
	\end{align}
where $N$ is the number of spins (electrons or holes) in the lead, $\langle S_z^{i} \rangle \equiv \langle \chi_i \vert S_z \vert \chi_i \rangle $ denotes the expectation value of the component of the spin operator along the quantization axis (`$z$-axis') when the $i$th particle is in the spin state $\chi$, and we assume a common quantization axis in both leads. In semiconductor physics, one is typically interested in the cases of electrons and heavy-holes. While the latter possess a total angular momentum of $3/2$ they may be described as an effective spin-$1/2$ system. To derive effective DSP dependent refilling rates, we consider the system to be in the (0,2) singlet state which is occupied at least once during a transport cycle. For a DSP in the left lead $p_L$ the probability of an electron entering the system by tunneling from the left lead to the left dot is given by $P_L(\sigma) = (1+ \sigma p_L)/2$, where $\sigma \in \lbrace 1,-1 \rbrace$ denotes the normalized magnetic spin quantum number. Similarly, the probability of an electron of spin $\sigma$ \textit{remaining} in the system after a time much larger than $1/\Gamma$ is $P_R(\sigma) = (1- \sigma p_R)/2$ for a DSP in the right lead $p_R$. Since the probability distributions are independent (the leads do not interact), the probability of achieving a configuration with an electron of spin $\sigma$ in the left dot and an electron of spin $\sigma'$ in the right dot is $P(\sigma, \sigma') = P_L(\sigma)P_R(\sigma')$. By multiplying the above probabilities with the dot-lead tunneling rate $\Gamma$, we obtain the effective refilling rates of the states in the (1,1) configuration,
	\begin{align}
		\label{eq:rates}
		\begin{split}
		&\Gamma_S = \Gamma_{T_0} = \frac{P(1,-1) + P(-1,1)}{2}  \Gamma = \frac{1+p_L p_R}{4} \Gamma, \\
		&\Gamma_{T_{\pm}} = P \left( \pm1, \pm 1 \right) \Gamma = \frac{(1\pm p_L)(1 \mp p_R)}{4} \Gamma.
		\end{split}
	\end{align}
As they are derived from probability considerations, the set of rates satisfies $ \Gamma_S + \Gamma_{T_0} + \Gamma_{T_+} + \Gamma_{T_-} = \Gamma$. Conventional PSB investigations take into account the special case of unpolarized leads, $p_L = p_R = 0$, for which all states are refilled equally with rate $\Gamma/4$. In  other extreme cases, one may work in a reduced state space when calculating the leakage current by choosing the applied magnetic field properly. An overview over some prominent special cases and the resulting refill rates is given in Table \ref{tab:polarizationcases}. In the following we look at the case of complete and equal DSPs in both leads in more detail, and derive an explicit analytical expression for the leakage current.
\begin{table}
		\begin{center}
		\rowcolors{2}{yellow!80!white!90}{yellow!50!white!40}
 			\begin{tabular}{|c| c|| c | c | c| c|} 
 			\hline
			Case & $(p_L,p_R)$ & $\Gamma_S/\Gamma$   & $\Gamma_{T_0}/\Gamma$& $\Gamma_{T_+}/\Gamma$ &  $\Gamma_{T_-}/\Gamma$\\ 
				 \hline \hline
 				I &$(1,1)$ & 1/2 & 1/2& 0 & 0 \\
 				II &$(-1,-1)$ & 1/2 & 1/2 & 0 & 0 \\
  				III &$(1,-1)$ & 0 & 0 & 1 & 0 \\ 
  				IV &$(-1,1)$ & 0 & 0 & 0 & 1 \\ 
  				V &$(0,0) $& 1/4 & 1/4 & 1/4 & 1/4 \\ 
				 \hline
			\end{tabular}
		\end{center}
	\caption{Special cases of spin polarization in the leads and the corresponding effective refill rates $\Gamma_i$ for the four (1,1) charge states. Cases I and II correspond to complete and parallel DSPs in the leads, characterized by $p_Lp_R =1$. Cases III and IV, on the other hand, correspond to complete and anti-parallel DSPs, $p_Lp_R = -1$. Case V is the standard case of unpolarized leads considered in conventional PSB investigations.}
	\label{tab:polarizationcases}
	\end{table}
\section{Leakage current}
\label{sec:leakage_current}
In this section we first review the basic model that describes transport through a DQD and allows us to calculate the leakage current. We then look at a special case of DSPs that allows us to reduce the state-space and obtain an exact analytical expression for the leakage current. Finally, we look at arbitrary DSPs and identify regions in which the g-tensor resonance found in Ref.~\cite{Mutter2020} is more pronounced compared to a setup with unpolarized leads.

To describe the transport quantitatively, we treat the DQD as an open system, described by the master equation,
	\begin{align} \label{eq:master_equation}
		i [H, \rho] =   \sum_k \Gamma_k \left( L_k \rho L_k^{\dagger} - \frac{1}{2} \lbrace L_k^{\dagger} L_k, \rho \rbrace \right) + D_{\text{rel}}[\rho],
	\end{align}
where the sum runs over all states in the (1,1) charge configuration, the curly brackets denote the anti-commutator, and we work in the steady state, $\dot{\rho} = 0$. The Hamiltonian $H$ is given in Eq.~\eqref{eq:H_DQD}, the rates $\Gamma_k$ are as given in Eq.~\eqref{eq:rates}, and $L_k = \vert k \rangle \langle S_{02} \vert$ are quantum jump operators in the framework of the Lindblad formalism. The left hand side of Eq.~\eqref{eq:master_equation} describes the unitary dynamics of the DQD system, while the right hand side contains dissipative processes. The first term on the right hand side of the equation describes the effective refilling processes, and the second term, $D_{\text{rel}}[\rho]$, models intrinsic relaxation processes, e.g., due to the spin-orbit interaction. Finally, the leakage current $I$ is determined by the probability of forming a singlet in the (0,2) charge configuration multiplied by the rate of this state to be emptied by a tunneling event to the right lead, $I = e\Gamma \langle S_{02} \vert \rho \vert S_{02} \rangle$.
\subsection{Ferromagnetic leads}
\label{subsec:ferromagnetic_leads}
We first consider the special case of parallel ferromagnetic leads, i.e., $p_L p_R = 1$ (Fig.~\figref[(a)]{polarized_leads_sketch}). If the magnetic field is chosen to be along one of the principal axes of the g-tensors, which are assumed to be diagonal in the same basis, $\mathbf{B} =B_k \hat{k}$, the effective magnetic fields in both dots are parallel, $\boldsymbol{\mathcal{B}}^L \parallel \boldsymbol{\mathcal{B}}^R$. Fixing the quantization axis along this direction, we may work in a reduced Hilbert space spanned by the states $\vert S_{02} \rangle$, $\vert S \rangle$ and $ \vert T_0 \rangle $. Moreover, we include isotropic relaxation processes, mediated by phonons, which occur when the thermal energy is larger than the Zeeman splitting, $k_BT > \vert g \vert B_k$, where $\vert g \vert $ is the largest component of the (diagonal) g-tensor. This type of relaxation does not lead to blocked states as every transition can occur in both directions. Working in the limit $\Gamma_{\text{rel}} \ll \Gamma, t$, one may then derive an effective relaxation rate from the unpolarized triplet to the singlet in the (1,1) charge configuration such that we may continue to work in the three-dimensional reduced state space.

\begin{figure}
		\includegraphics[width=0.95\linewidth]{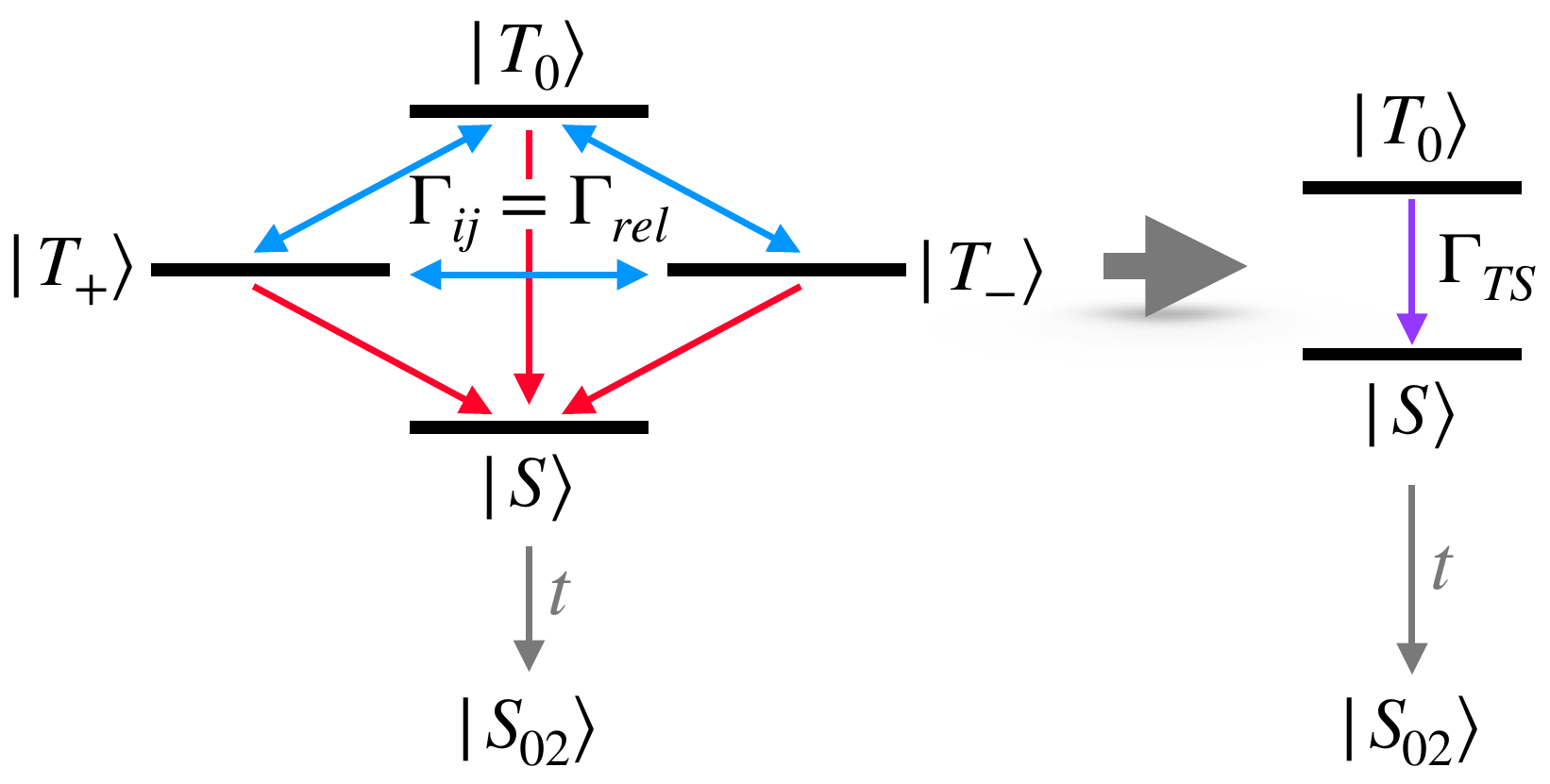}
		\caption{Visualization of the effective rate approach. In the full five-state model the unpolarized triplet state $|T_0\rangle$ transitions with a rate $\Gamma_{\text{rel}}$ to another state (to any particular of the three remaining states with a rate $\Gamma_{\text{rel}}/3$). Blue arrows indicate transitions within the triplet subspace, while red arrows indicate one-way transitions from the triplets to the singlet in the (1,1) confuguration which will subsequently tunnel into the (0,2) configuration without further relaxation. In the three state model one may combine all these rates into a single effective rate $\Gamma_{\text{TS}} $ from the unpolarized triplet to the singlet (purple arrow, Eq. \eqref{eq:expectedrate}).}
		\label{fig:relaxationratescheme}
	\end{figure}

\begin{figure}
		\includegraphics[width=0.95\linewidth]{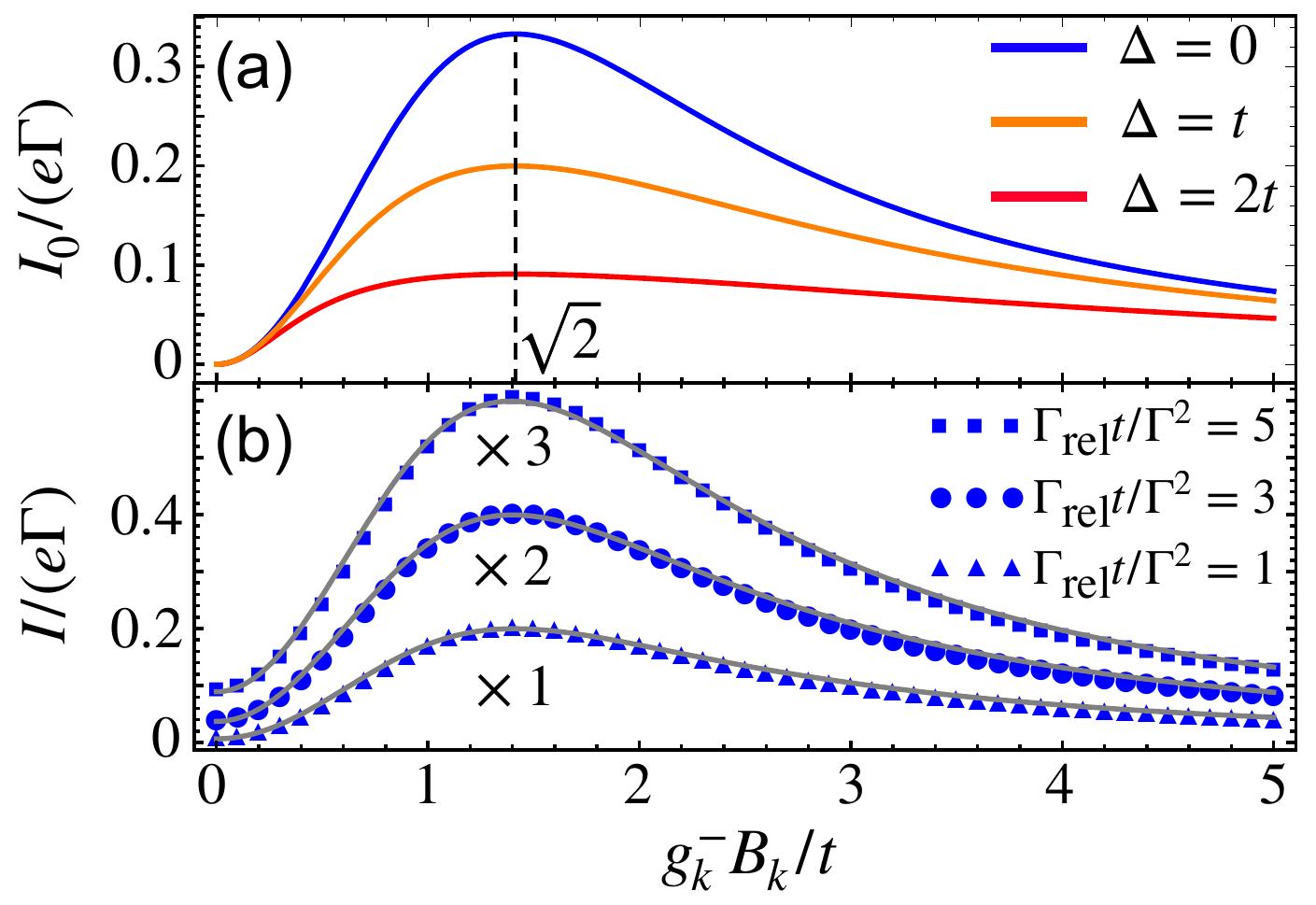}
		\caption{Leakage current for $p_L p_R = 1$. (a) The exact analytical magnetotransport curve $I_0(B_k)$ as given by Eq. \eqref{eq:analyticalcurrent}. It shows a distinct maximum at $B_k^* = \sqrt{2}t/g_k^-$ independent of the detuning $\Delta$. (b) The leakage current including high-temperature relaxation processes at zero detuning, $\Delta = 0$. The zero field value is finite and agrees well with the analytical result, Eq. \eqref{eq:Irelzero}. For the sake of clarity the curves and numerical data points are scaled by a factor as indicated in the figure. The dot-lead tunneling rate was set to $\Gamma = 0.1t$ for both plots. }
		\label{fig:analyticalcurrent}
	\end{figure}

When a singlet is formed, the electron in the left dot will tunnel to the right dot and from there to the lead before the singlet can relax into any of the triplets. At zero field, the triplet $\vert T_0 \rangle$ is a blocked state. If the system is refilled into the $\vert T_0 \rangle$ state, the blockade can only be lifted by relaxation chains ending in the singlet state $\vert S \rangle$, which will transition coherently into the (0,2) configuration without further relaxation. Assuming equal rates among all (1,1) states, the probability for the process $\vert T_0 \rangle \rightarrow \vert S \rangle$ along paths including $n$ transitions (there are $2^{n-1}$ such paths) is given by $P(n) = 2^{n-1}/3^n$, satisfying $\sum_{n=1}^{\infty}  P(n)= 1$ via a geometric series. Consequently, the expected number of transitions needed to reach the singlet is $\langle n \rangle = 3$. Since all occupied triplets transition with a rate  $\Gamma_{\text{rel}}$, the total rate along a path with $n$ transitions is $\Gamma_n = \Gamma_{\text{rel}}/n$. Therefore, the expected value for the rate connecting the unpolarized triplet to the singlet is
	\begin{align}
	\label{eq:expectedrate}
	 	\Gamma_{\text{TS}} \equiv \langle \Gamma_n \rangle = \Gamma_{\text{rel}}  \sum_{n=1}^{\infty} \frac{2^{n-1}}{n3^n} \approx 0.55 \;\Gamma_{\text{rel}}.
	\end{align}
We treat this as an effective triplet-singlet relaxation rate in the reduced space spanned by the states $\lbrace \vert S_{02} \rangle , \vert S \rangle, \vert T_0 \rangle \rbrace$. The presence of the polarized triplet states $\vert T_{\pm} \rangle$ results in a reduced relaxation rate of the unpolarized triplet. While $\vert T_0 \rangle$ is emptied with a rate $\Gamma_{\text{rel}}$ in the full five-state model, it is effectively only emptied with a rate $ \Gamma_{\text{TS}}  \approx 0.55 \;\Gamma_{\text{rel}}$ in the reduced three-state model (Fig.~\ref{fig:relaxationratescheme}).
As a result, the dissipator in Eq.~\eqref{eq:master_equation} has the form
	\begin{align}
	\label{eq:master_equation_high_T}
	\begin{split}	
		D_{\text{rel}}[\rho] = \Gamma_{\text{TS}}  \left( \vert S \rangle \langle T_0 \vert \rho\vert T_0 \rangle \langle S \vert - \frac{1}{2} \left\lbrace\vert T_0 \rangle \langle T_0 \vert, \rho  \right\rbrace \right).
	\end{split}
	\end{align}
The leakage current $I = e\Gamma  \langle S_{02} \vert \rho \vert S_{02} \rangle$ is obtained by exactly solving the master equation in Eq.~\eqref{eq:master_equation} including the dissipative term in Eq. \eqref{eq:master_equation_high_T} with the normalization constraint $\text{Tr} \rho = 1$. It is instructive, however, to first solve the system in the absence of relaxation processes, $\Gamma_{\text{TS}}  = 0$,
where one finds
	\begin{align}
	\label{eq:analyticalcurrent}
		I_0 = \frac{2e \Gamma t^2}{2t^2 + (g_k^-B_k)^2 + 4t^4/(g_k^-B_k)^2 + 4 \Delta^2 + \Gamma^2},
	\end{align}
where $g_k^- = g_k^L - g_k^R$. The magnetotransport curve $I_0(B_k)$ possesses characteristic maxima at $B_k^* = \pm \sqrt{2} t/g_k^-$ where the current takes the value $I_0^{\text{max}} = 2e \Gamma t^2 /(6t^2 + 4 \Delta^2 + \Gamma^2)$ (Fig. \figref[(a)]{analyticalcurrent}). The occurrence of a maximum can be understood qualitatively. On the one hand, the energy separation between the singlet-triplet hybridized states grows with increased magnetic field, decreasing the current. On the other hand, the rate with which the unpolarized triplet may transition to the singlet via the hybridized states as a result of different out-of-plane g-factors in the dots is increased for larger magnetic field strengths. These counter-acting effects lead to an optimal value of the magnetic field. When $ \Gamma_{\text{TS}} \neq 0$ the analytical expression for the current is modified. However, as we can see from Fig.~\ref{fig:analyticalcurrent} it is changed qualitatively only at zero field, where the current becomes finite and takes the value
	\begin{align}
	\label{eq:Irelzero}
		I \left(B_k =0 \right) = \frac{3}{5} e \Gamma \left(2 +\frac{\Delta^2}{t^2} +\frac{\Gamma^2}{4t^2} + \frac{\Gamma}{2 \Gamma_{\text{TS}} }   \right)^{-1}.
	\end{align}
The factor of $3/5$ is due to the temporary population of the polarized triplet states $\vert T_{\pm} \rangle $ which are blocked. Consequently, only three out of five states contribute to the transport, causing the reduction. Fig. \figref[(b)]{analyticalcurrent} shows a comparison between the exact analytical solution to the effective three-state model and the numerical magnetotransport curve for the full five-state model. For the numerics we use the full isotropic relaxation term $D_{\text{rel}}[\rho] = -\Gamma_{\text{rel}} \rho + \sum_{j,d} \sigma_j^d \rho \sigma_j^d /6$, where $j \in \lbrace x,y,z \rbrace$ runs over all three Pauli matrices and $d$ labels the two dots. We see good agreement between the exact analytical solution and the numerical results. The zero-field value depends on the relaxation rate, and thus allows the determination of the latter from magnetotransport measurements. Additionally, the  position of the maximum is robust against the effects of isotropic relaxation, and its measurement allows us to determine the difference in out-of-plane g-factors. Together with the g-tensor resonance found in Ref.~\cite{Mutter2020}, $B_k^* = t/\sqrt{g_k^L g_k^R}$, the g-factors along $\hat{k}$ may thus be completely determined without any prior knowledge necessary.
	
Clearly, the current will vanish in the presence of low temperature spin-relaxation processes, since there are one-way transitions to the blocked spin ground state $\vert T_- \rangle$ (for $g_z^{L/R} >0$). As these relaxation processes are expected to be present at cryogenic temperatures used in quantum information technology, one must apply a magnetic in-plane field (i.e., one that is perpendicular to the spin quantization axis) to couple the polarized triplets to the unpolarized triplet and the singlet to obtain a non-zero current. The next section is devoted to investigating the case of general polarizations in this setup.
\subsection{Arbitrary polarizations}
\label{subsec:arbitrary_polrizations}
\begin{figure}
		\includegraphics[width=0.97\linewidth]{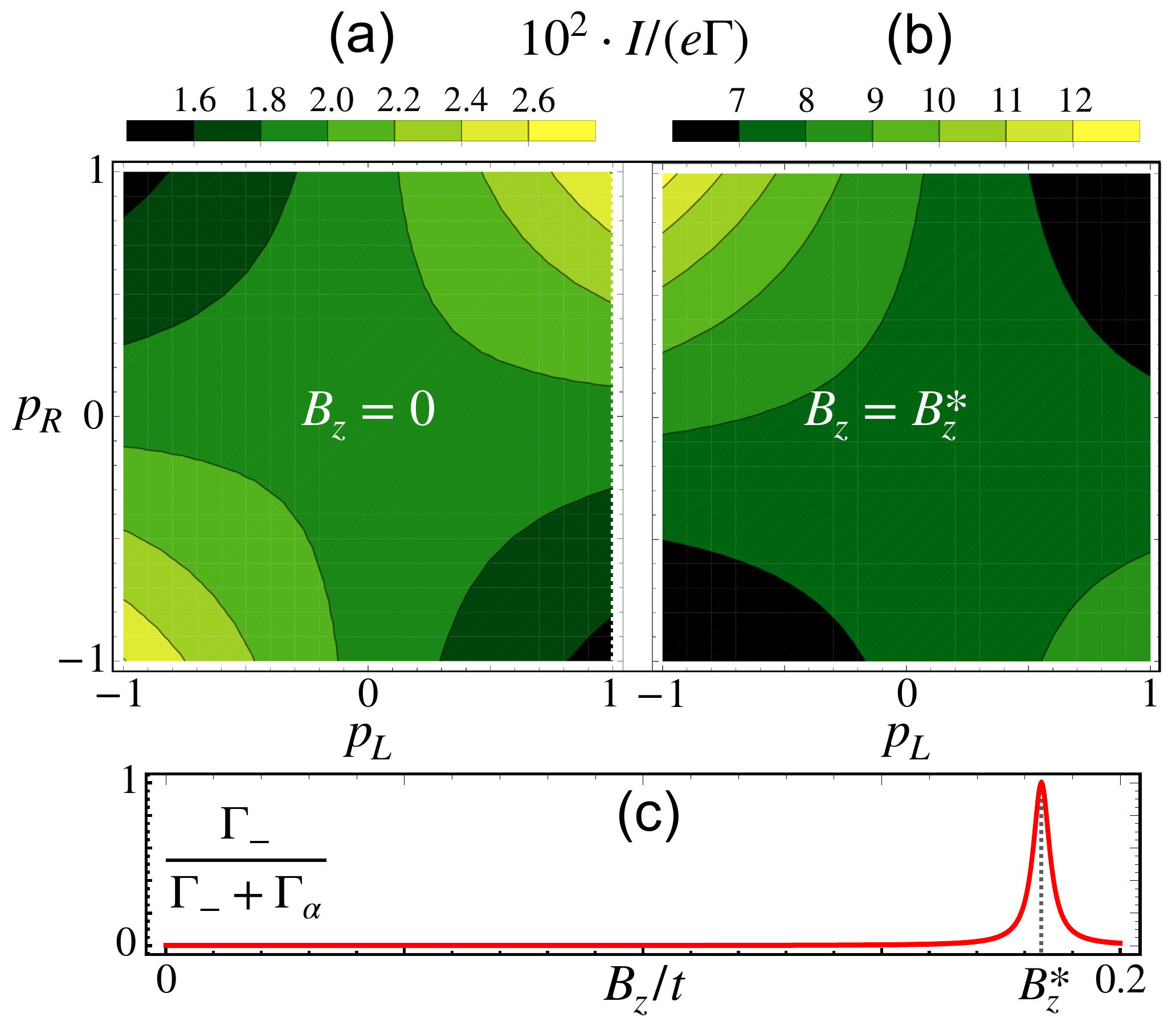}
		\caption{Leakage current $I(p_L,p_R)$ as a function of the DSPs in the left and right leads. (a) shows the case of vanishing out-of-plane magnetic field, $B_z = 0$, and (b) fixes the magnetic field at the g-tensor resonance of the system, $B_z^* = t/\sqrt{g_z^L g_z^R}$. The regions in the $p_L$-$p_R$-plane for which the current is highest are exchanged as the magnetic field is increased, an observation that can be explained on grounds of energy considerations (see text). (c) The normalized ratio of the rate of the polarized triplet $\vert T_- \rangle$ transitioning to the (0,2) configuration, $\Gamma_-$, and the rate of the hybridized states $\vert \alpha_{0, \pm} \rangle$  transitioning to the (0,2) configuration, $\Gamma_{\alpha}$,  as a function of the out-of-plane magnetic field. The parameter values used for all plots are $\Delta = 0$, $g_z^L = 5.5$, $g_z^R = 5.4$, $g_x^L = 0.4$, $g_x^R = 0.3$, $\Gamma_{\text{rel}} = 0.001t$, $\Gamma_R = 0.01t$ and $B_x = 0.1t$.}
		\label{fig:contourplotscomparison}
	\end{figure}

For the case of arbitrary spin-polarization in the leads we solve the steady-state master equation in Eq.~\eqref{eq:master_equation} in the full PSB Hilbert space $\mathcal{H}~=~\text{span} \lbrace \vert S_{02} \rangle , \vert S \rangle, \vert T_0 \rangle, \vert T_+ \rangle, \vert T_- \rangle \rbrace$ numerically. We include low temperature relaxation processes with rate $\Gamma_{\text{rel}}/2$ described by the dissipator
	\begin{align}
		D_{\text{rel}}[\rho] = \frac{\Gamma_{\text{rel}}}{2}\sum_{d \in \lbrace L,R \rbrace} \left(S_-^d \rho S_+^d - \frac{1}{2} \left\lbrace S_+^d S_-^d, \rho  \right\rbrace \right),
	\end{align}
where $S_{\pm}^d= S^d_x \pm i S^d_y$ are the spin ladder operators in dot $d$.
Additionally, we now allow for a magnetic field with both in- and out-of-plane components, $\mathbf{B} = (B_x,0,B_z)$. We aim to explore the effect of spin-polarized leads on the leakage current in the g-tensor resonance setup of Ref.~\cite{Mutter2020}. For this we work at zero detuning, $\Delta = 0$ and small in-plane magnetic fields, $(g_x^L + g_x^R) B_x \ll t$. In this limit the dominant Hamiltonian is diagonal in the two polarized triplets and three hybridized states $\vert \alpha_{0, \pm} \rangle$ mixing the singlets and the unpolarized triplet. Transitions between these states are due to different g-tensors in the dots and yield a non-zero leakage current that depends on the DSP in the leads, which we display in Fig. \ref{fig:contourplotscomparison}. 
	
At zero out-of-plane magnetic field, Fig.~\figref[(a)]{contourplotscomparison}, the configurations with $p_Lp_R = 1$ produce the highest current because half of the time a singlet state is formed in the (1,1) charge configuration which can directly transition into the spin singlet in the (0,2) charge configuration (Fig.~\figref[(a)]{polarized_leads_sketch}). The configurations with $p_Lp_R = -1$ show a reduced current because now a polarized triplet is formed when the DQD is refilled (Fig.~\figref[(b)]{polarized_leads_sketch}). These states must first transition to the (1,1) singlet (with rate $\sim g_x^- B_x$) before the blockade is lifted.  At the g-tensor resonance, i.e., for $B_z= B_z^* =t/\sqrt{g_z^L g_z^R}$, the effect of the DSPs in the leads is rather different (Fig.~\figref[(b)]{contourplotscomparison}). The configuration $(p_L, p_R) = (-1,1)$ shows the highest current since only the ground state triplet $\vert T_- \rangle$ is refilled and can transition resonantly and unhindered by the relaxation processes to the (0,2) charge configuration via hybridized states containing a superposition of $\vert T_0 \rangle$, $\vert S \rangle $ and $\vert S_{02} \rangle$. The cases with $p_L p_R =1$ show the lowest current as the rate $\Gamma_{\alpha}$ of transitioning to the (0,2) singlet from the hybridized states $\vert \alpha_{0, \pm} \rangle$ is much smaller than the rate of transitioning to the (0,2) singlet from the polarized triplets (Fig.~\figref[(c)]{contourplotscomparison}). The asymmetry between the the cases $(p_L, p_R) = (-1,1)$ and $(p_L, p_R) = (1,-1)$ is induced by the low temperature relaxation processes since the triplets are split by the Zeeman energy for non-zero magnetic fields. While the triplet $\vert T_+ \rangle$ can also transition resonantly to the (0,2) configuration, it is at risk of relaxing into the unpolarized triplet or the singlet, thereby slowing down the transport. This behaviour may be used in experiments to amplify the magnetotransport signal. Since the current is reduced for $p_Lp_R=-1$ at zero field but enhanced at $B^*$, the g-tensor resonance is more pronounced in systems where the leads show complete but opposite DSPs. 

Remarkably, we find a giant magnetoresistance when the magnetic field is tuned to the g-tensor resonance $B_z^*$. Fixing the left lead to be completely polarized with $p_L=-1$, we find a large sensibility of the current on the DSP in the right lead. Defining the resistance coefficient~\cite{Baibich1988,Binasch1989}
    \begin{align}
    \label{eq:magnetoresistance}
        \delta(p^1_R,p^2_R) = \frac{R(p^1_R) - R(p^2_R)}{R(p^2_R)} = \frac{I(p^2_R) - I(p^1_R)}{I(p^1_R)},
    \end{align}
where $R(p_R) \propto 1/I(p_R)$ is the resistance when the right lead has a DSP $p_R$, we find values of $\delta(-1,1) $ exceeding 100 percent. Hence, a tip consisting of a ferromagnetic lead coupled to a DQD held at the g-tensor resonance may be used to read out the DSP in the target lead.

We remark that in this work we do not consider the effect of the hyperfine interaction of the electron or hole spin with the host nuclei spins, which is a small effect in many materials of interest, e.g., for heavy-holes in germanium and may be neglected completely when working with isotopically purified materials. Also, cotunneling processes are not taken into account. While both effects have been shown to affect the leakage current~\cite{Jouravlev2006, Golovach2004, Vorontsov2008, Qassemi2009, Coish2011}, we expect no qualitative change in our results when they are included in the model. Indeed, the two additional terms in the Hamiltonian would yield an increased current which is most pronounced at zero field, while the effects of site-dependent g-tensors and non-trivial DSPs in the leads result in magnetotransport features at finite fields.

Furthermore, to introduce the transport formalism and investigate the effects of spin-polarized leads together with site-dependent g-tensors in a clear and disentangled fashion, the SOI has only been considered indirectly as one mechanism for spin relaxation so far. We now turn to a more complete description of the DQD system by explicitly including the SOI, which is ubiquitous in solid state systems and of great interest for qubit gate manipulation in spin based quantum information technology~\cite{Mutter2020cavitycontrol, Mutter2021natural}.

\section{Spin-orbit interaction}
\label{sec:SOI}
The SOI plays an integral role in many materials of interest such as valence band states in germanium, and it is responsible for a number of effects that may be harnessed in qubit engineering. In this section we analyze the effect of the SOI on the leakage current and how the DSPs in the leads affect characteristic features known from conventional PSB investigations.

For systems where the SOI preserves time reversal symmetry, the spin-flip tunneling terms induced by the SOI are described by the general Hamiltonian,
	\begin{align}
	\label{eq:H_SOI}
		H_{\text{SO}} = i \mathbf{t}_{\text{SO}} \cdot \vert \mathbf{T} \rangle \langle S_{02} \vert + \text{H.c.},
	\end{align}
where $\mathbf{t}_{\text{SO}}  = (t_x,t_y,t_z)$ is the spin-orbit vector of the system and 
	\begin{align}
		\vert \mathbf{T} \rangle =\begin{pmatrix}
			\vert T_x \rangle \\
			 \vert T_y \rangle  \\
			\vert  T_z \rangle
		\end{pmatrix} 
		=  \frac{1}{\sqrt{2}}\begin{pmatrix}
			\vert T_- \rangle - \vert T_+ \rangle \\
			i \left( \vert T_- \rangle + \vert T_+ \rangle \right) \\
			\sqrt{2}  \vert  T_0 \rangle
		\end{pmatrix}
	\end{align}
is a vector containing triplet states in a combination such that $\mathbf{t}_{\text{SO}} $ transforms as a real vector under coordinate transformations ~\cite{Danon2009}. An example of non time reversal symmetric SOI is provdided by heavy-holes in semiconductors where parts of the effective SOI are induced by a magnetic field.

\begin{figure}
		\includegraphics[width=0.98\linewidth]{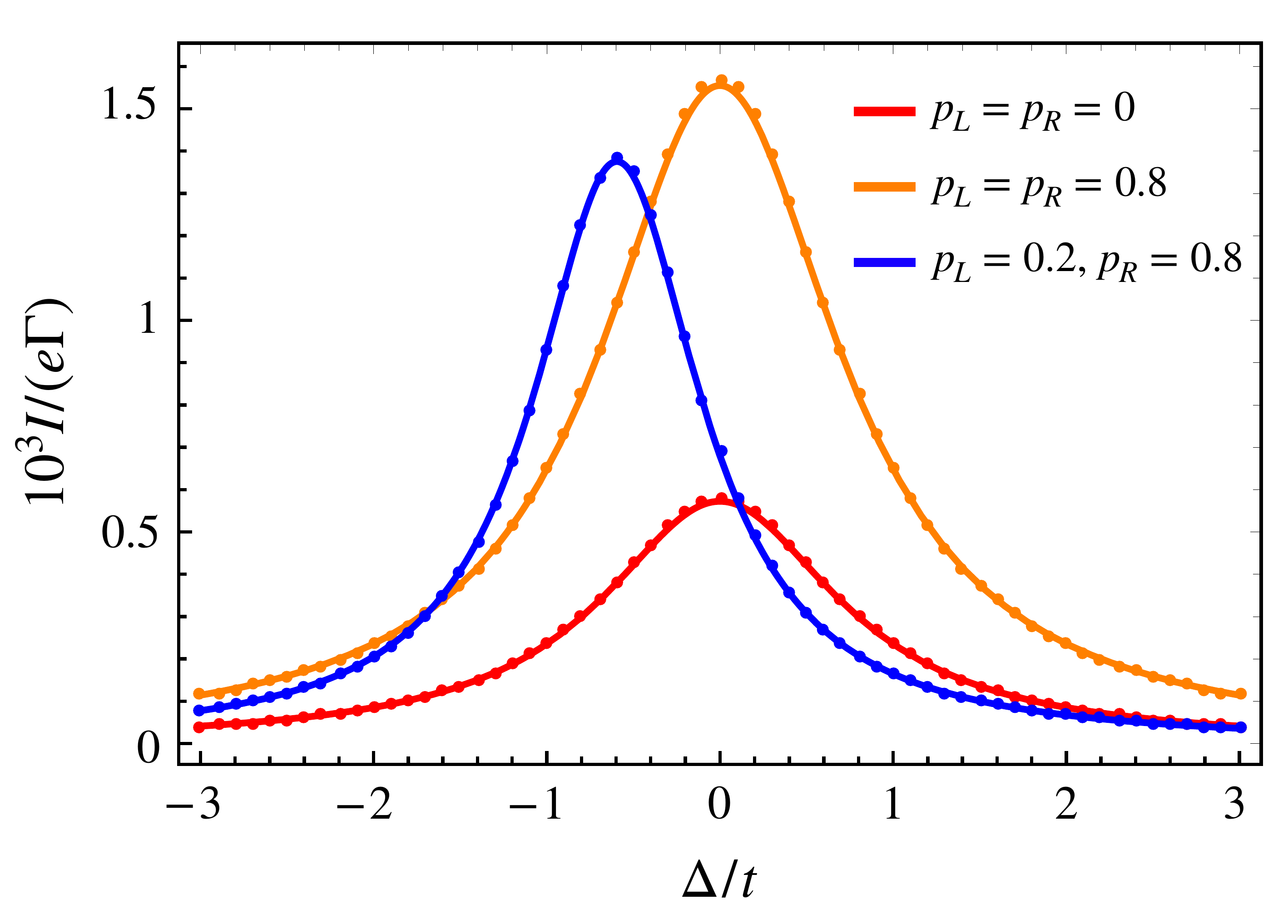}
		\caption{Leakage current in the presence of spin-flip tunneling processes induced by the SOI. We plot the current $I$ as a function of the detuning $\Delta$ and compare the numerical results (dots) to the analytical expressions (solid lines, Eq.~\eqref{eq:analytical_current_arbitrary_DSP}). The current is is enhanced for non-zero lead polarizations if $p_L p_R >0$ (orange curve). When the DSPs differ in the two leads, the current becomes asymmetric in the detuning (blue curve). We set $B_z = 0.3t$, $g_z^L = 5.5$, $g_z^R = 4.5$ and $t_x = t_y = t_z = 0.01t$.}
		\label{fig:current_SOI}
	\end{figure}

In the limit $B_z g_z^-, t_{\text{SO}} \ll t$, where $t_{\text{SO}} = \vert \mathbf{t}_{\text{SO}} \vert$, one may compute the current for an applied out-of-plane field of magnitude $B_z$ according to the formula,
	\begin{align}
		\frac{I}{e \Gamma} = \left( \sum_i \frac{\Gamma_i}{\Gamma^{\text{decay}}_i} \right)^{-1},
	\end{align}
where the sum runs over all blocked states, the refill rates $\Gamma_i$ are as given in Eq.~\eqref{eq:rates} and the decay rates of the blocked state $\vert i \rangle$ transitioning to the (0,2) singlet $\Gamma_i^{\text{decay}}$ are computed using first-order perturbation theory. We find
	\begin{align}
	\label{eq:analytical_current_arbitrary_DSP}
	\begin{split}
		& \frac{e \Gamma}{I} = 
		\frac{e \Gamma}{I_+} +  p_Lp_R \frac{e \Gamma}{I_-}
		+ \frac{p_R-p_L}{2}  \frac{ (B_z g_z^+)^2 -4t^2}{B_z g_z^+ (t_x^2+t_y^2)} \Delta,\\
		&\frac{e \Gamma}{I_{\pm}} = \left(\frac{t}{B_z g_z^-} \right)^2 \pm \frac{2 t^4 \Lambda(B_z, \Delta)}{(t_x^2+t_y^2) (B_z g_z^+)^2} , \\
		& \Lambda(B_z, \Delta) =  \left( \left(\frac{B_z g_z^+ }{2t} \right)^2 - 1 \right)^2 + \left( \frac{\Delta B_z g^+_z }{2 t^2} \right)^2,
	\end{split}
	\end{align}
where $I_+$ is the current for non-polarized leads, $I_-$ contains the effect of spin-polarization ($p_L, \; p_R \neq 0$) and the last term in the first line of Eq.~\eqref{eq:analytical_current_arbitrary_DSP} introduces an asymmetry in the detuning and the magnetic field when the DSP in the leads are not equal ($p_L \neq p_R$, Fig.~\ref{fig:current_SOI}). The term $I_-$ increases the current when $p_L p_R >0$, i.e., when the leads are both predominantly filled with one spin state. The maximum of the current as a function of the detuning at fixed magnetic field is found to be
	\begin{align}
	\label{eq:max_position_detuning}
		\Delta_{\text{max}} = \frac{p_L - p_R}{1-p_Lp_R}  \Delta_r , \;  \Delta_r = \frac{(B_z g_z^+)^2 - 4t^2}{2B_z g_z^+}.
	\end{align}
Note that this shift is independent of the SOI in the system and thus provides a clear signature of different DSPs in the leads which can be picked up in experiment.  As $p_Lp_R \leq 1$, the sign of $\Delta_{\text{max}}$ is determined by the sign of $(p_L - p_R) \Delta_r$, $\Delta_r$ being the resonant detuning where the hybridized singlet energies due to the tunnel coupling align with the Zeeman energies of the polarized triplets. At $B_z > 0$ the low-energy triplet $\vert T_- \rangle$ is energetically favourable over the ground state singlet for $\Delta_r >0$. If $p_L < p_R$ a spin down state is more likely to enter the system than to leave it. To compensate for this polarization bias, the (0,2) configuration must be energetically favourable and hence we find the maximum of the current at negative detunings, $\Delta_{\text{max}} < 0$. On the other hand, if the triplet is not favourable ($\Delta_r < 0$) in an otherwise equal situation, the (1,1) configuration must be favoured to increase spin down refill events leading to $\Delta_{\text{max}} > 0$. If there is no bias in polarization across the system, the triplet is equally likely to be refilled as emptied and no compensating detuning between (1,1) and (0,2) is required, and the current has a maximum at zero detuning as for the case of unpolarized leads.

We note that when the magnetic field has an in-plane component, we observe the same behaviour as in Sec.~\ref{subsec:arbitrary_polrizations} but with the generalized g-tensor resonance from Ref.~\cite{Mutter2020}. In the simplest case, when the spin-orbit vector is parallel to one of the principal axes of the g-tensor and the dominant magnetic field is applied along this direction, $\mathbf{B} = B_k \hat{k}$, one has $B^*_k = \sqrt{t^2+t_{\text{SO}}^2}/\sqrt{g_k^L g_k^R}$. As before, the current may be enhanced at this point by tuning the leads close to the configuration with $p_Lp_R = -1$.

\section{Conclusion}
\label{sec:conclusion}
We show that the DSPs in the leads connected to a DQD influence the leakage current. By working in a reduced state-space for the case of equally oriented ferromagnetic leads, we obtain an exact analytical expression for the leakage current, which contains information about the relaxation rate via its zero field value and the g-tensor components in the dots via its position of the maximum. The latter allows for a full determination of the g-tensors when combined with the resonance proposed in Ref.~\cite{Mutter2020}, which we show to be more pronounced  compared to the standard case of unpolarized leads when the DSPs are tuned close to the point $(p_L,p_R) =(-1,1)$. Moreover, when the system is operated at the g-tensor resonance, we observe a giant magnetoresistance exceeding 100 percent, suggesting the possibility of using a lead coupled to a DQD with site-dependent g-tensors as a read out tip. Finally, we incorporate the effects of the SOI and find that the above results still hold when working with a generalized g-tensor resonance. Moreover, we demonstrate that one may obtain information about the DSPs by recoding the leakage current for various detunings. Our results pave the way for an exact determination of the g-tensors in a DQD system and the DSPs in the leads by magnetotransport measurements alone.

\appendix
\section{Dot-lead spin-flip tunneling}
\label{appx:dot-lead_spin-flip_tunneling}
One may lift the assumption that the tunneling processes to and from the leads are spin-conserving. If the probability for a spin-flip tunneling event from/to lead $d \in \lbrace L,R \rbrace$ is given by $r_d < 1/2$, then the altered probabilities of an electron or hole of normalized spin $\sigma \in \lbrace -1,1 \rbrace$ entering the system from the left lead ($d=L$) or remaining in the right dot ($d=R$) are,
	\begin{align}
	\label{eq:spin-flip_probabilities}
		\tilde{P}_d(\sigma) = P_d (\sigma) \left( 1 - 2 r_d \right) +  r_d,
	\end{align}
where $P_L(\sigma) = (1+ \sigma p_L)/2$, $P_R(\sigma) = (1- \sigma p_R)/2$ for a degree of spin polarization in the left (right) lead $p_L$ ($p_R$). The effective dot-lead tunneling rates may then be calculated by inserting the probabilities in Eq.~\eqref{eq:spin-flip_probabilities} into Eq.~\eqref{eq:rates} of the main text. In the case where the spin exchange with the leads becomes completely random, $r= 1/2$, one has $\tilde{P}_d(\sigma) = 1/2$, and thus $\Gamma_i = 1/4$ for all states $\vert i \rangle$.
\bibliographystyle{apsrev4-2}
\bibliography{PSBleadsliterature}
\end{document}